\documentclass[prl,
superscriptaddress,
 amsmath,amssymb,
 aps,
twocolumn
]{revtex4-2}

\usepackage{setspace}
\usepackage{amsmath}
\let\oldAA\AA
\renewcommand{\AA}{\text{\normalfont\oldAA}}
\usepackage{xcolor}
\usepackage{multirow}

\usepackage{booktabs}
\usepackage{lipsum}
\usepackage{graphicx}
\usepackage{dcolumn}
\usepackage{bm}
\usepackage{amsmath}
\usepackage{amssymb}
\usepackage{dutchcal}
\usepackage{float}
\begin{document}
\title{Electrical resistivity, thermal conductivity, and viscosity of Fe-H alloys at Earth’s core conditions}
\author{Cong Liu}
\email{Corresponding author: cliu10@carnegiescience.edu}
\affiliation{Extreme Materials Initiative, Earth and Planets Laboratory, Carnegie Institution for Science, 5241 Broad Branch Road NW, Washington, District of Columbia 20015, USA}

\author{R.~E. Cohen}
\email{Corresponding author: rcohen@carnegiescience.edu}
\affiliation{Extreme Materials Initiative, Earth and Planets Laboratory, Carnegie Institution for Science, 5241 Broad Branch Road NW, Washington, District of Columbia 20015, USA}

\begin{abstract}
The transport properties (electrical resistivity, thermal conductivity, and viscosity) of iron-hydrogen alloys are of great significance in the stability and evolution of planetary magnetic fields. Here, we investigate the thermal transport properties of iron doped with varying hydrogen content as functions of pressure (P) and temperature (T) for the top and bottom of Earth’s outer  core and beyond, corresponding to pressures of about 130 to 380 GPa and temperatures of 4000 to 7000 K. Using first-principles density functional theory  molecular dynamic simulations (FPMD), we verify that crystalline FeH$_x$ is superionic with H diffusing freely. We find a  low frequency viscosity of 10-11 $mPa\cdot s$ for liquid Fe-H alloys at Earth's outer core conditions. We find resistivity saturation with increasing temperatures in liquid Fe-H alloy at core pressures. The effect of H on electrical and thermal transport we find is small, so that the exact H content of the core is not needed. The primary effect of H is on the equation of state, decreasing the density at constant P and T.  We find the Lorenz number is smaller than the ideal value, and obtain for X(H)= 0.20, or 0.45 wt\% H , thermal conductivity $\kappa$ of $\sim$105 and $\sim$190 $Wm^{-1}K^{-1}$ at conditions near the core-mantle and inner-outer core boundary, respectively.

\end{abstract}
\maketitle

\section{Introduction}
The Earth's core, primarily an iron-nickel alloy, has 5-10\% lower density than pure iron at core conditions, suggesting the presence of lighter elements like hydrogen, carbon, oxygen, silicon, and sulfur\cite{poirierLightElementsEarth1994,hiroseLightElementsEarth2021a}. These elements may affect the core's physical and chemical properties, such as viscosity, electrical resistivity, and thermal conductivity, which are crucial for understanding the core's thermal evolution, geomagnetic field dynamics, and overall behavior of Earth's interior\cite{dewijsViscosityLiquidIron1998,pozzoThermalElectricalConductivity2012,konopkovaDirectMeasurementThermal2016}.

Challenging experimental studies have explored electrical and thermal transport in iron and its alloys as functions of pressure and temperature \cite{gomiElectricalResistivityThermal2015,ohtaExperimentalDeterminationElectrical2016,konopkovaDirectMeasurementThermal2016,zhangReconciliationExperimentsTheory2020,edmundFeFeSiPhaseDiagram2022}, but studying the effect of hydrogen on transport is exceedingly difficult experimentally because controlling and measuring small or variable hydrogen content in situ in the diamond anvil cell is very hard.  Even for pure iron, there are significant variations in measured transport properties. For instance, direct measurements\cite{konopkovaDirectMeasurementThermal2016} of thermal conductivity under conditions close to those of Earth's core have yielded low values, such as 46 Wm$^{-1}$K$^{-1}$. These results support the conventional thermal dynamo theory and are consistent with a geodynamo driven by thermal convection throughout Earth's history. Other indirect measurements \cite{ohtaExperimentalDeterminationElectrical2016} have given higher values (226 Wm$^{-1}$K$^{-1}$), as have  estimates of thermal conductivity from electrical resistivity measurements in conjunction with the Wiedemann-Franz law, $\kappa=LT/\rho$, where thermal conductivity $\kappa$ and electrical resistivity $\rho$ are associated by temperature T and Lorenz number $L$, which is often estimated as the ideal low-temperature value, $L_0$.

A number of density functional theory (DFT) based first-principles molecular dynamics \cite{zhangReconciliationExperimentsTheory2020,zhangThermalConductivityFeSi2022,pozzoSaturationElectricalResistivity2016,pozzoThermalElectricalConductivity2014a}and linear response computations \cite{shacohen2011} have also estimated iron transport properties .   High values of thermal conductivity were computed using molecular dynamics \cite{pozzoThermalElectricalConductivity2012}, and earlier by first-principles linear response lattice dynamics. \cite{shacohen2011} These high values are inconsistent with the thermal convection mechanism in the core, suggesting the need for an alternative mechanism to explain the geodynamo\cite{badroEarlyGeodynamoDriven2016,orourkePoweringEarthDynamo2016}.Recent theoretical calculations suggest a moderate thermal conductivity of approximately 100$\pm$10 Wm$^{-1}$K$^{-1}$ of hexagonal close-packed ($hcp$) Fe under core-mantle boundary (CMB) conditions with both electron-phonon ($e$-$ph$) and electron-electron ($e$-$e$) scattering contributions \cite{xuThermalConductivityElectrical2018}.

Previous research indicates that introducing light elements can significantly influence the transport properties, elastic wave velocities, and viscosity of Earth's interior\cite{greff-lefftzCoremantleCouplingViscoelastic1995,ichikawaAtomicTransportProperty2015,liAtomicTransportProperties2021,wagleResistivitySaturationLiquid2019,inoueResistivitySaturationHcp2020,zhangThermalConductivityFeSi2022}. Even a small amount of silicon doping greatly effects the transport properties of iron at the Earth's core conditions\cite{zhangThermalConductivityFeSi2022}. Given that hydrogen is the most abundant element in the universe, it plays a pivotal role in the Earth's mantle and core. Understanding the effects of an iron-hydrogen alloy under core conditions is particularly intriguing. The superionic behavior\cite{houSuperionicIronOxide2021,wangStrongShearSoftening2021,zhangDirectHHeChemical2022,heSuperionicIronAlloys2022a,yangIronHydrideEarth2022,parkElectrideFormationHCPIron2024} suggests that hydrogen, when incorporated into iron under extreme pressure and temperature, could lead to unique transport properties. This includes potentially altered electrical and thermal conductivities, which are critical for modeling the core's thermal evolution and geomagnetic field dynamics. Additionally, the viscosity of the core, which influences convective processes and the geodynamo mechanism, could be markedly affected by the presence of hydrogen.

\section{Methods}

To compute the pressure and perform configuration sampling at given volumes and compositions, first-principles molecular dynamics (FPMD) simulations were carried out using Quantum ESPRESSO \cite{giannozziQUANTUMESPRESSOModular2009a}. The Perdew–Burke–Ernzerhof (PBE) generalized gradient approximation was employed for the exchange-correlation functional \cite{perdewGeneralizedGradientApproximation1996a}, along with scalar-relativistic Garrity–Bennett–Rabe–Vanderbilt (GBRV) ultrasoft pseudopotentials \cite{GARRITY2014446} for Fe and H. A plane wave cutoff energy of 40 Ry was applied. Simulations were performed in the NVT ensemble with ionic temperatures regulated by a stochastic-velocity rescaling thermostat, and electronic temperatures were included via the Fermi–Dirac smearing function. A periodic $hcp$ supercell containing 128 iron atoms was doped with various hydrogen concentrations (0, 1, 2, 4, 8, 16, and 32 H atoms, corresponding to 0–0.45 wt\%) at interstitial sites. The validity of the MD simulations was confirmed by testing parameters such as simulation time, time step, and pseudopotentials (Fig. S1 in the supplemental material \cite{sm_supplemental}). At $V=1071$ $\AA^3$, simulations were conducted at 4000 K to represent the solid/fluid iron phase and at 6000 K for the fluid phase. At $V=905$ $\AA^3$, simulations were performed at 6000 K to represent the solid/fluid phase and at 7000 K for the fluid phase. The $c/a$ ratio was fixed at 1.615, with additional tests performed at a $c/a$ ratio of 1.7 to model $hcp$ iron under Earth's core conditions \cite{steinle2001}. The Brillouin zone was sampled at the $\Gamma$-point, with an FPMD time step of 1 fs and total simulation durations exceeding 10 ps to ensure equilibrium. Tests were performed with a time step of 0.25 and no significant changes were found. Bands with occupations greater than $10^{-5}$ were included in the calculations, and macroscopic quantities such as temperature and pressure were averaged over 3000 equilibrium steps. Additionally, $bcc$ Fe-H configurations were tested using supercell of 128 and 160 atoms at $V=905$ $\AA^3$ and 6000 K, employing the same MD simulation parameters. Details of these tests are provided in Fig. S2 of the supplemental material \cite{sm_supplemental}.  We also performed some FPMD simulations with the Vienna ab initio simulation package (VASP) to cross check our MD results in the Supplemental Material \cite{sm_supplemental} (see also references \cite{kresseEfficientIterativeSchemes1996,blochlProjectorAugmentedwaveMethod1994,perdewGeneralizedGradientApproximation1996a,gonzalez-cataldoInitioDeterminationIron2023,steinle2001} therein).

The self-diffusion coefficient $D$ is derived from the mean square displacement $r^2(t)$ according to the following equations:
\begin{equation}
    D=r^2(t)/6t
\end{equation}
\begin{equation}
    r ^2 \left ( t \right ) =\left \langle \left [ r_i \left ( t + t_0 \right ) - r_i \left ( t_0 \right ) \right ] ^2  \right \rangle_{i,t_0}
\end{equation}

From the fluid equilibrium MD trajectories, we calculated the viscosity \(\eta(t)\) of the liquid Fe-H alloy using the linear-response Green–Kubo formula. This involved integrating the autocorrelation function (SACF) of the off-diagonal stress tensor \(\sigma\), \(C(t)\), from 0 to \(t\). The integration was efficiently performed using fast Fourier transforms and the convolution theorem, enabling averaging over time steps.

\begin{equation}
    C(t)=\frac{V}{k_B T}\left\langle \sigma(t) \sigma(t+t_0) \right \rangle_{t_0}
\end{equation}
\begin{equation}
    \eta(t)= \int_{0}^{t} C(t_0) dt_0
\end{equation}
The final viscosity $\eta$ is taken for $t\to\infty$. For SACF, we used a cutoff of 2 ps and then fitted the data with two exponentials. This reproduces well the long time behavior of C(t) beyond the ballistic region.\cite{liAtomicTransportProperties2021}
 \begin{equation}
     C(t)=G_\infty \left[ (1-\alpha)sech^2(t/\tau_1)+\alpha e^{-t/\tau_2}\right]
 \end{equation}

We obtain the frequency dependent viscosity $\eta(\omega)$ and shear modulus $G(\omega)$ for the fluid from the Fourier transform of C(t).
\begin{equation}
    \eta(\omega)=\int_{0}^{\infty} C(t)e^{-i\omega t}dt
\end{equation}
\begin{equation}
    G(\omega)=i\omega\eta(\omega)
\end{equation}

where $\sigma$ represents the three independent components of the off-diagonal stress tensor:$\sigma_{xy}$,$\sigma_{yz}$,$\sigma_{zx}$, and $V, k_B, T$ are the cell volume, the Boltzmann constant, and the temperature, respectively.
 
The Stokes–Einstein relation between viscosity and diffusivity is given by:
\begin{equation}
    \frac{D\eta}{\rho^{1/3}k_BT}=a,
    \label{eq:se}
\end{equation}
where $\rho=N/V$ is the number density, and $a$ is the Stokes-Einstein dimensionless parameter. For binary system of Fe-H alloy, we firstly average the diffusion constants of Fe and H \cite{liuStokesEinsteinRelationBinary2024}:
\begin{equation}
    D_{avg}=\frac{N_{Fe}D_{Fe}+N_HD_H}{N_{Fe}+D_h}.
\end{equation}

\begin{figure*}[htp]
\centering
\includegraphics[width=0.7\textwidth]{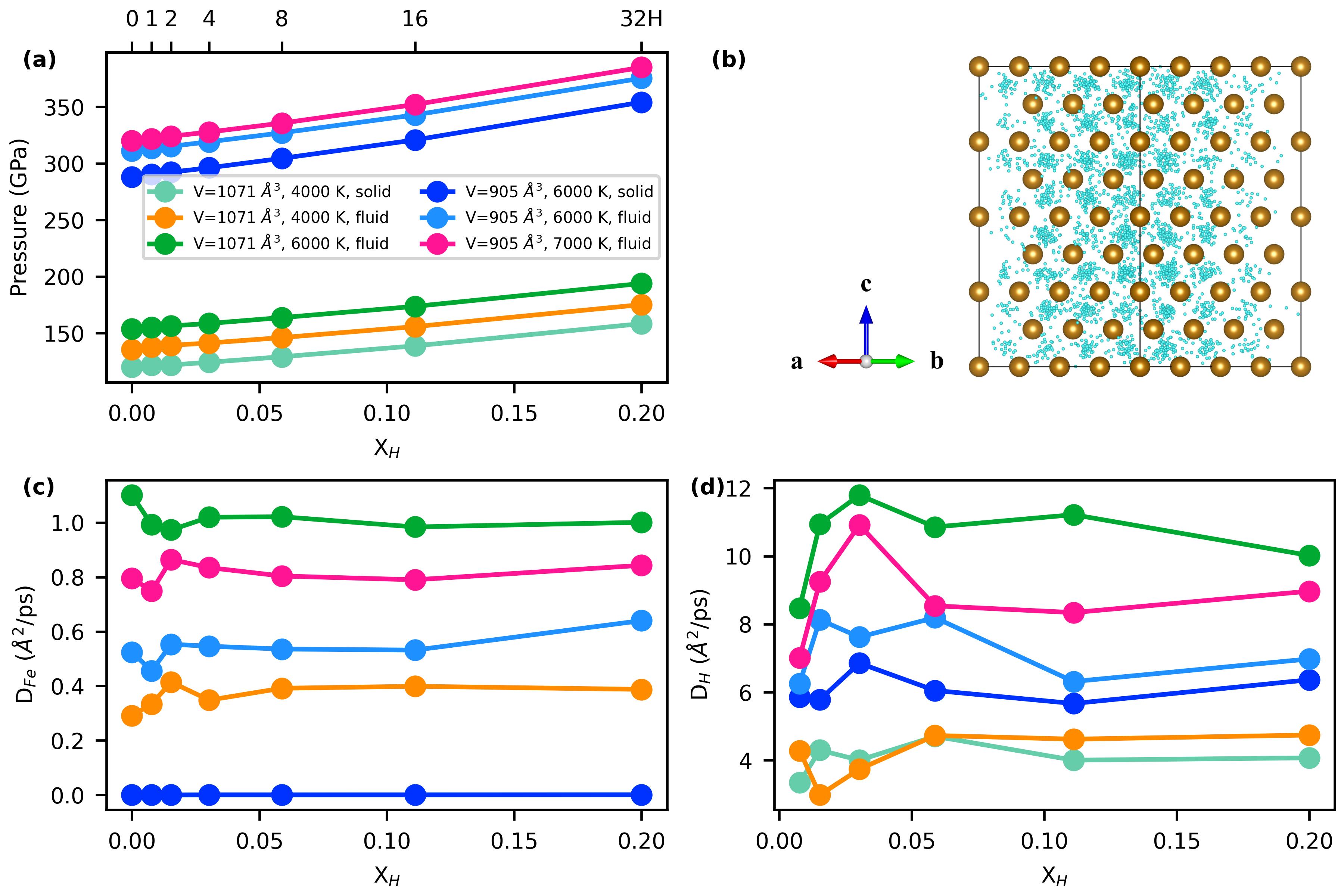}
\caption{\textbf{First-principles molecular dynamics simulations for iron-hydrogen alloy at Earth's core conditions.} \textbf{a}~Effects of hydrogen content on pressure at constant volume and temperature.  Six curves are divided into two sets: solid at 4000 K (light green), fluid at 4000 K (orange), fluid at 6000 K (green) at larger volume (V=1071 $\AA^3$ per supercell of 128 iron atoms); and solid at 6000 K (blue), fluid at 6000 K (light blue), and fluid at 7000 K (magenta) at smaller volume (V=905 $\AA^3$ per same). \textbf{b}~MD trajectory of solid Fe$_{128}$H$_{32}$ at 6000 K and V=905 $\AA^3$. Brown and cyan atoms represent iron and hydrogen, respectively. Chemical diffusivity of \textbf{c} Fe and \textbf{d } H  are derived from the slope of $\langle r^2(t)\rangle$. The top axes labels indicate the number of H in the 128 Fe supercell.}
\label{fig:md}
\end{figure*}

We employed a modified Spin-polarized Relativistic Korringa-Kohn-Rostoker (SPR-KKR) package \cite{ebertCalculatingCondensedMatter2011,minarCorrelationEffectsTransition2011} to perform transport calculations for Fe-H alloys, using randomly sampled snapshots from AIMD simulations. Three configurations from FPMD simulations of the Fe-H alloy were analyzed to compute the electron-phonon (\(e\)-\(ph\)) scattering contributions to electrical and thermal conductivity using density functional theory (DFT). Additionally, both electron-phonon and electron-electron (\(e\)-\(ph/e\)-\(e\)) scattering contributions were calculated using dynamical mean field theory (DMFT). In liquids, resistivity arises from scattering caused by liquid disorder, as liquids lack well-defined phonons\cite{xuThermalConductivityElectrical2018,zhangReconciliationExperimentsTheory2020}.

For each configuration, calculations began with a self-consistent-field (SCF) step using density functional theory (DFT). A maximum angular momentum \( l_{\text{max}} = 3 \) (\(nl = 4\)) was employed, with 27 k-points for the supercell, corresponding to \(27 \times (128\)-160) k-points for the primitive cell. We also tested \( l_{\text{max}} = 4 \) in the KKR method and observed similar results (Fig. S3 in the supplemental material \cite{sm_supplemental}). Following the SCF step, transport calculations were conducted using the converged potentials, applying the Kubo-Greenwood method with \( l_{\text{max}} = 3 \), 64 k-points, and energy values within \([-5 k_B T, 5 k_B T]\) around the chemical potential. The computed values for electrical resistivity (\(\rho\), the inverse of electrical conductivity, \(1/\sigma\)) and thermal conductivity (\(\kappa\)) are given as: 

\begin{equation}
    \sigma=\mathcal{L}_{11}
\end{equation}
\begin{equation}
    \kappa=\frac{1}{eT}\left(\mathcal{L}_{22}-\frac{\mathcal{L}_{12}^2}{\mathcal{L}_{11}}\right)
\end{equation}
and $\mathcal{L}_{ij}$ is given by:
\begin{equation}
    \mathcal{L}_{i j}=(-1)^{(i+j)} \int d \varepsilon \sigma_{\mu \nu}(\varepsilon-\mu)^{(i+j-2)}\left(-\frac{\delta f_T(\varepsilon)}{\delta \varepsilon}\right)
\end{equation}
At a given temperature T, $f_T(\varepsilon)$ represents the Fermi-Dirac distribution function. The energy-dependent conductivity is given by the Kubo expression :
\begin{equation}
    \sigma_{\mu \nu} \propto \operatorname{Tr}\left\langle\hat{j}_\mu \operatorname{Im} G^{+} \hat{j}_\nu \operatorname{Im} G^{+}\right\rangle
\end{equation}
where $\hat{j}_\mu$ is the current density operator, and $\operatorname{Im} G^{+}$ is the imaginary part of retarded Green function. For transport calculations, We used a maximum angular momentum $l_{max}$ of 4 and energies window of $\pm$5$k_B$T around the chemical potential with a uniform k-point sampling of 4$\times$4$\times$4 for the supercell.

We also calculate the electrical resistivity due to electron-electron scattering by fully relativistic KKR-DMFT implemented in the SPR-KKR package. We used the SPTF (spin-polarized T matrix + FLEX (fluctuation exchange)) impurity solver with a Hubbard parameter U=4.0 eV and the J=0.943 eV. The analytical continuations of the bath Green’s function and electronic self-energy are done using averages over Padé approximates. For SCF step, we used a maximum angular momentum $l_{max}$ of 3 with 2 k-points for supercell. Subsequently, transport calculations were performed using the converged potentials obtained from the SCF step, employing the Kubo-Greenwood method with a maximum angular momentum $l_{max}$ of 3 on 64 k-points for the supercell and energies [-5 k$_B$T,5 k$_B$T] around the chemical potential. Tests show that our results are converged with respect to all parameters.

\begin{figure*}[ht]
\centering
\includegraphics[width=0.7\textwidth]{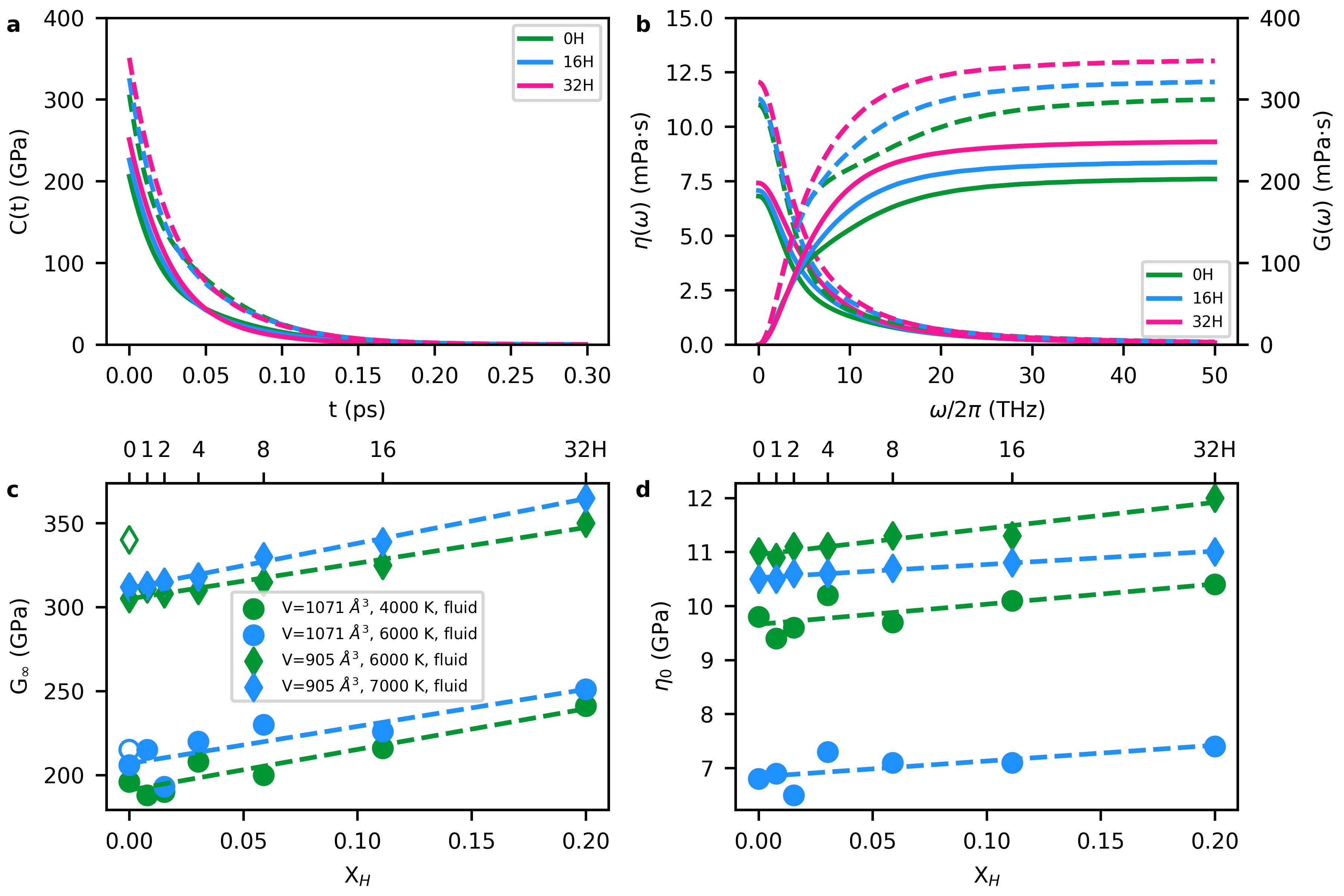}
\caption{\textbf{Viscosity of liquid iron-hydrogen alloy at Earth's core conditions.} \textbf{a}~Stress autocorrelation functions (SACFs) of liquid iron-hydrogen alloy at 6000 K  and V=1071 (solid), 905 (dash) $\AA^3$  by doping 0H (green, pure iron), 16H (blue) and 32H (red) from fitting. See  Fig.\ref{fig:md}a for the corresponding pressures. {b}~Frequency-dependent viscosity ($\eta(\omega)$, lines decreasing to zero, corresponding to left y-axis) and shear modulus ($G(\omega$, lines increasing, corresponding to right y-axis) of liquid iron at 6000 K  and V=1071 (solid), 905 (dash) $\AA^3$ with various hydrogen doping content. Only the real parts of $\eta(\omega)$ and $G(\omega$) are shown. \textbf{c}~ Infinite frequency shear modulus $G_{\infty}$ and \textbf{d}~zero frequency viscosity $\eta_0$ (green, lower T and blue, higher T) at V=1071 (dot) and 905 (diamond) $\AA^3$. The lines are linear fits to the points. The open diamond and circle represent results under the same conditions as the solid symbols from previous studies\cite{xianViscoelasticityLiquidIron2019}.}
\label{fig:viscosity}
\end{figure*}

\section{Results}

 In our fixed cell simulations, we observed a linear increase in thermal pressure with the addition of hydrogen doping (Fig.\ref{fig:md}a). This trend indicates that hydrogen incorporation into the iron lattice contributes significantly to the internal pressure. Hydrogen increases the pressure at constant volume, or increases the volume and decreases the density at constant pressure. In the solid phases, we found that hydrogen atoms can move freely within the iron $hcp$ lattice (Fig.\ref{fig:md} b), which is the so-called superionic state\cite{cavazzoniSuperionicMetallicStates1999,liuMultipleSuperionicStates2019}.

Despite the significant impact of hydrogen content on the equation of state (i.e. pressure at given volume), its influence on the self-diffusivity of iron is minor(Fig.\ref{fig:md} c, d). The self-diffusivity of iron $D_{Fe}$ remains almost unchanged with varying hydrogen content. However, the self-diffusivity of hydrogen $D_H$ shows some fluctuations, especially at low hydrogen doping levels, due to the lattice deformation induced by the small amounts of hydrogen. We also found that iron fluid at 6000 K at a volume of 1071 $\AA^3$ (P=153-193 GPa) has a larger self-diffusion coefficient than iron fluid at 6000 and 7000 K in a volume of 905 $\AA^3$  (P=311-384 GPa), indicating that the diffusivity is strongly dependent on both temperature and pressure. Higher temperatures increase the kinetic energy of atoms, resulting in more vigorous vibrations and greater lattice deformation. Larger volumes at lower pressure allow atoms to move easier, enhancing diffusivity. 

\begin{figure}[ht]
\centering
\includegraphics[width=0.45\textwidth]{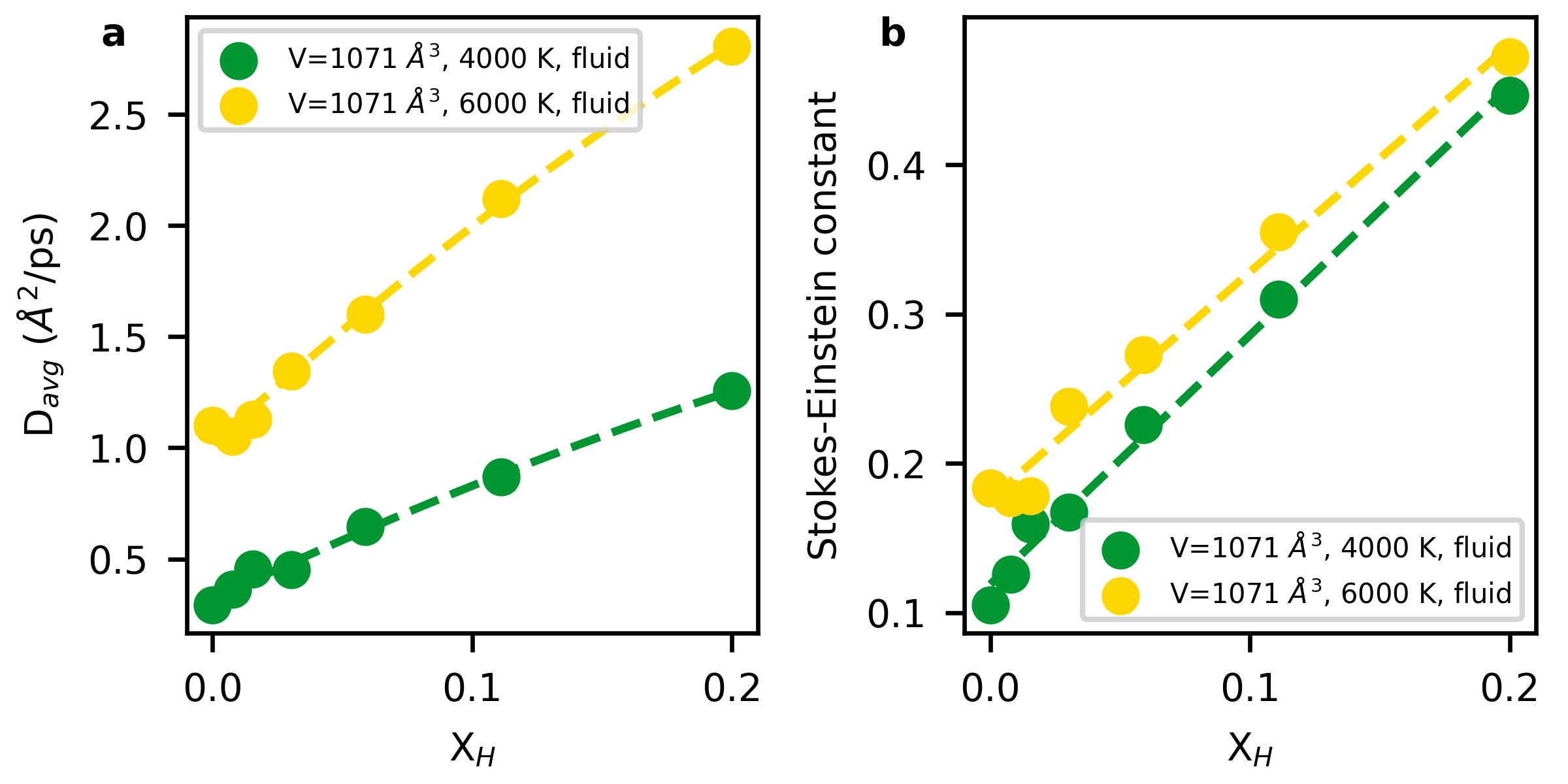}
\caption{\textbf{Stokes-Einstein relation at a fixed volume of 1071 $\AA^3$  with increasing doping H content for 4000 K fluid phase and 6000 K fluid phase.} \textbf{a} Composition averaged self-diffusion coefficient $D_{avg}$ and \textbf{b} Stokes-Einstein dimensionless constant $a$ as a function of the doping H.}
\label{fig:SE}
\end{figure}

\begin{figure*}[ht]
\centering
\includegraphics[width=0.7\textwidth]{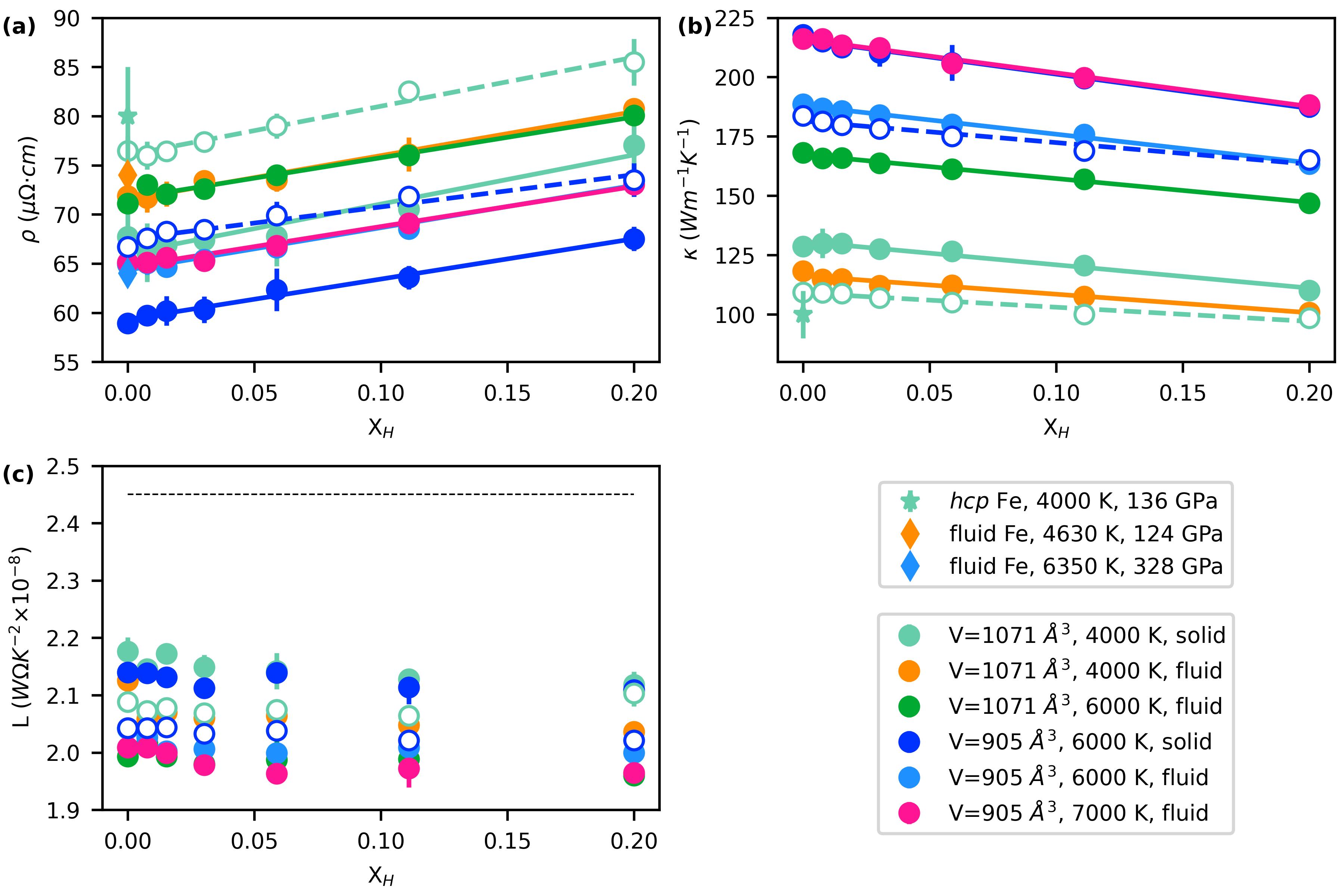}
\caption{\textbf{Electrical transport properties of Fe-H alloy at the Earth's core conditions.}  \textbf{a} Electrical resistivity and \textbf{b} thermal conductivity evolution in two fixed volumes at different temperatures with doping H content. See  Fig.\ref{fig:md}a for the corresponding pressures. Six curves are divided into two sets: solid phase at 4000 K (light green), fluid phase at 4000 K (orange), fluid phase at 6000 K (green) in a larger volume (V=1071 $\AA^3$); and solid phase at 6000 K (blue), fluid phase at 6000 K (light blue), fluid phase at 7000 K (magenta) in a smaller volume (V=905 $\AA^3$). The error bars represent standard deviation from configurations sampling. DFT/DMFT results are marked with solid/open circles. We fit linear solid/dash curves for these discrete DFT/DMFT values that share the same volume, temperature, and state. The light green stars and orange/light blue diamonds are from computations in  Ref\cite{zhangReconciliationExperimentsTheory2020}and Ref\cite{pozzoTransportPropertiesLiquid2013a}. The ideal Lorenz number (L$_0$=2.44$\times$10$^{-8}$ W$\Omega$K$^{-2}$) is shown as the dotted horizontal line at the top.}
\label{fig:transport}
\end{figure*}

We obtained the shear viscosity from the the off-diagonal stress tensor $\sigma$ autocorrelation function (SACF), $C(t)$, which decays rapidly when $t$ is small \cite{liAtomicTransportProperties2021}. The fitted SACF converge to $\sim$ zero at t=0.15 ps (Fig. \ref{fig:viscosity}a) and show little dependence on H content. We obtain the frequency dependent viscosity $\eta(\omega)$ and shear modulus $G(\omega)$ for the fluid from the Fourier transform of C(t). The viscosity, $\eta(\omega)$, decreases monotonically with frequency, and $G(\omega)$ increases from zero and plateaus at high frequencies (Fig. \ref{fig:viscosity} b). From these numerical integration of the original molecular dynamics (MD) data, we derive the viscosity and shear modulus, corresponding to the zero-frequency viscosity $\eta_0$ and the infinite-frequency shear modulus $G_\infty$, respectively. Our calculated \( G_\infty \) values for liquid iron under outer and inner core conditions are slightly lower than those reported in previous DFT studies. We find that the viscosity of the Fe-H alloy at 6000 K and a volume of 1071 $\AA^3$ does not vary significantly with hydrogen doping levels of 0, 16, and 32 H atoms (corresponding to P=153, 173, and 193 GPa). However, the shear modulus increases with higher hydrogen doping content. The shear modulus and viscosity increase a lot at higher pressure when we vary the volumes at 6000 K. Additionally, we compared the viscosity and shear modulus at different temperatures (Fig. \ref{fig:viscosity} (c-d)). With increasing hydrogen content, the shear modulus increases significantly due to the rise in internal pressure caused by the hydrogen. The shear modulus of the Fe-H alloy slightly decreases with higher temperatures. In contrast, the viscosity is minimally affected by hydrogen doping but decreases with rising temperatures. We obtain a viscosity of $\sim$10.0 $mPa\cdot s$ for the Fe-H alloy under Earth's outer core conditions, which is $\sim$ 20\% higher than previous results\cite{liAtomicTransportProperties2021}.

Besides the exact linear-response approach, the viscosity of liquid iron has also been approximately determined from diffusivity using the Stokes–Einstein relation under the high-pressure and high-temperature conditions of Earth's outer core. The chemical diffusivity of Fe and H increase linearly with hydrogen content (Fig. \ref{fig:SE}), leading to a linearly increasing Stokes–Einstein dimensionless parameter $a$ (Eq. \ref{eq:se}). For a fixed density, high temperatures enhance the average diffusivity of the Fe-H alloy but decrease the inverse $k_B T$. Thus we find that the Stokes–Einstein constant varies only slightly between 4000 and 6000 K, which verifying the validity of the Stokes–Einstein relation . The behavior of average diffusivity is also  significantly dependent on the H content and lead to a sharply increasing Stokes–Einstein constant by increasing H content.

\begin{figure*}[htp]
\centering
\includegraphics[width=0.8\textwidth]{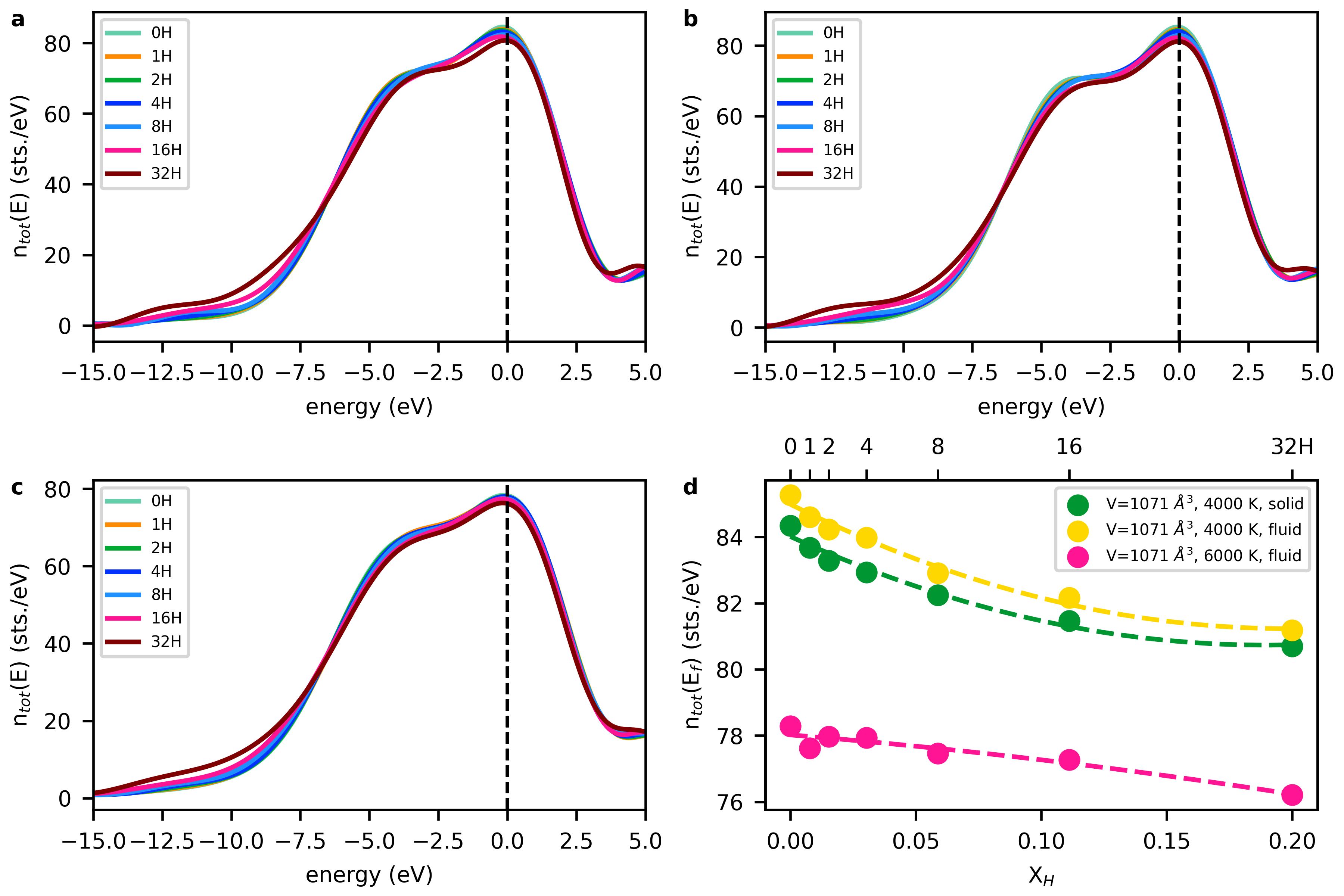}
\caption{\textbf{Total density of states (DOS) at the the fixed volume of 1071 $\AA^3$  with increasing doping H content} for \textbf{a} 4000 K solid phase, \textbf{b} 4000 K fluid phase, and \textbf{c} 6000 K fluid phase. The energy windows are rescaled with E$_{Fermi}$=0 by dash black line. \textbf{d} Intensity of DOS at Fermi surface with increasing doping H content.}
\label{fig:dos}
\end{figure*}

Other than atomic transport properties calculations, we furthermore carry out simulations for electrical resistivity and thermal conductivity of Fe-H alloy, which are crucial for our understanding of the Earth's core evolution. We computed the electrical resistivity and thermal conductivity using the Green-Kubo method for MD snapshots (Fig.\ref{fig:transport}). For pure iron (0H), we obtained electrical resistivity values of 68$\pm$3 and 76$\pm$0.2 $\mu\Omega\cdot cm$ for the $hcp$ phase using DFT and DMFT calculations at outer core conditions (130 GPa and 4000 K). The corresponding thermal conductivity values are 127$\pm$5 and 110$\pm$1 $Wm^{-1}K^{-1}$, consistent with previous FPMD+DFT/DMFT results\cite{zhangReconciliationExperimentsTheory2020}. The computed electrical resistivity of liquid iron is higher by $\sim$4 $\mu\Omega\cdot cm$ and the thermal conductivity lower by $\sim$ 10 $Wm^{-1}K^{-1}$ compared to the $hcp$ phase iron at 135 GPa and 4000 K. Heating iron to 6000 K and 153 GPa resulted in saturation of electrical resistivity in liquid iron, and the thermal conductivity increased to 130$\pm$4 $Wm^{-1}K^{-1}$, approximately 20\% lower than previous DFT results\cite{dekokerElectricalResistivityThermal2012}. Matthiessen’s rule is expected to be broken in the saturation region when the resistivity approaches the Ioffe-Regel limit, as this limit represents the minimum mean free path, essentially the nearest-neighbor distance\cite{gurvitchIoffeRegelCriterionResistivity1981}. For strongly correlated systems of Fe-H alloy at core conditions, resistivity can significantly exceed the Ioffe-Regel limit, corresponding to an extremely short mean free path.

For solid iron at the top of the inner core, we obtained electrical resistivity values of 59$\pm$0.2 and 67$\pm$0.2 $\mu\Omega\cdot cm$ for the $hcp$ phase using DFT and DMFT calculations. The corresponding thermal conductivity values are 220$\pm$5 and 183$\pm$1 $Wm^{-1}K^{-1}$. Upon heating from 6000 K to 7000 K at $\sim$350 GPa, we again observed saturation of electrical resistivity in liquid iron, which was also reported in previous light elements impurity studies\cite{pozzoSaturationElectricalResistivity2016,gomiElectricalResistivitySubstitutionally2016,wagleResistivitySaturationLiquid2019,inoueResistivitySaturationHcp2020}.  From the solid phase to the liquid phase at 6000 K, the electrical resistivity increases by $\sim$6 $\mu\Omega\cdot cm$, whereas the thermal conductivity decreases by $\sim$16\%. In fluid iron, thermal conductivity increases with rising temperature.

The total electronic density of states (DOS) of these molecular dynamics (MD) configurations at outer core conditions reveals that the DOS reaches its maximum at the Fermi level (Fig. \ref{fig:dos}). However, the DOS at the Fermi level decreases with increasing hydrogen (H) content, which in turn induces higher electrical resistivity in the $hcp$ Fe-H alloy. The error bars displayed are larger in solid iron phases compared to liquid phases due to the anisotropy  of the solid (in other words our ``error bars" include the variations in direction.). The electrical resistivity saturates at high temperatures regardless of H content, although impurities do increase the resistivity values (Fig. \ref{fig:transport}a). At outermost core conditions, a hydrogen content of 20 mol\% (0.45 wt\%) increases the electrical resistivity by $\sim$ 10\% compared to pure iron, according to density functional theory (DFT) calculations (solid orange line). Our dynamical mean-field theory (DMFT) results also show that electrical resistivity increases linearly with H content and are moderately higher than those obtained from DFT.

\begin{figure*}[ht]
\centering
\includegraphics[width=0.7\textwidth]{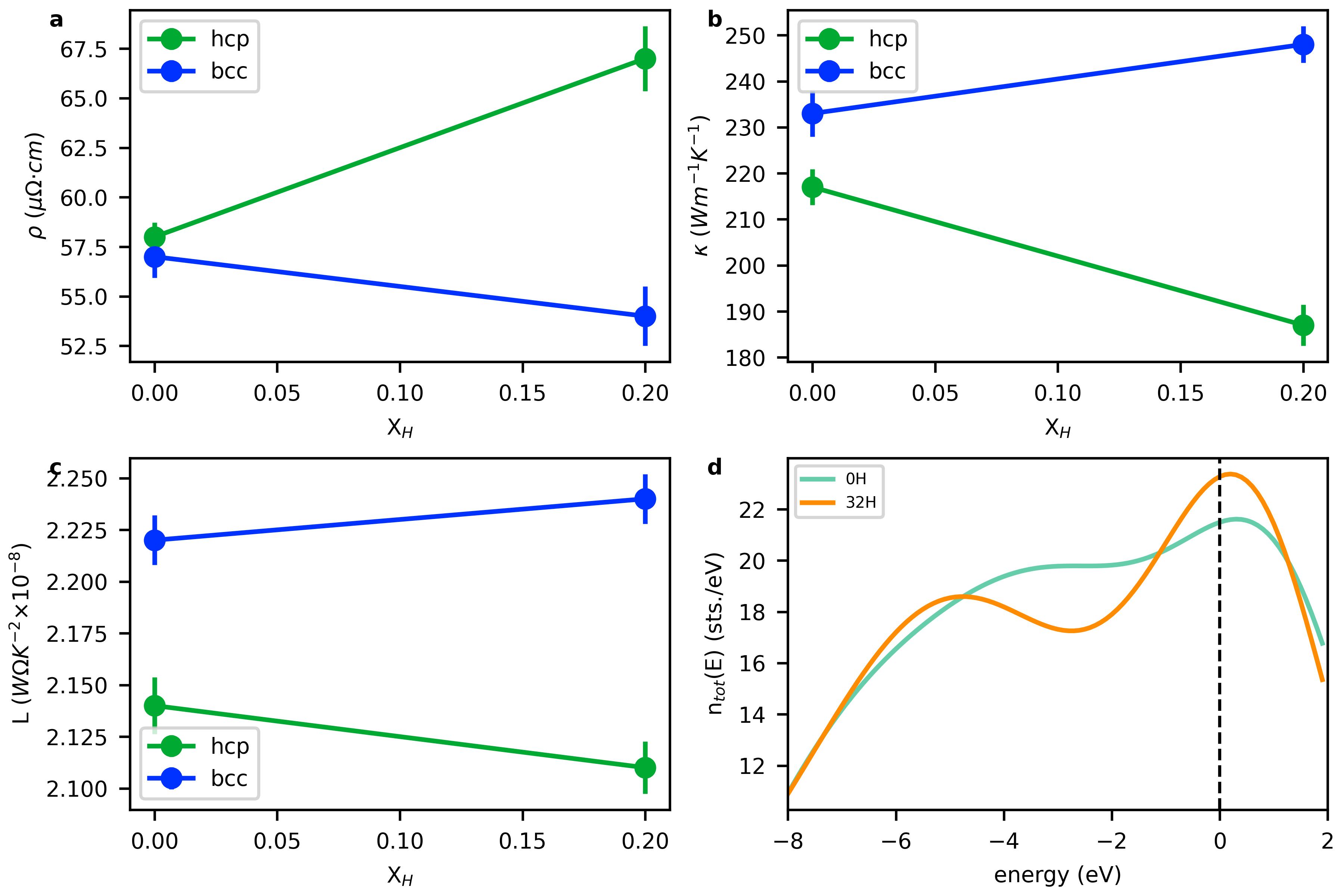}
\caption{\textbf{Electrical transport properties of different solid phase (bcc and hcp) Fe-H alloy at 6000 K in the fixed volume of 905 $\AA^3$  in KKR.}  \textbf{a} Electrical resistivity, \textbf{b} thermal conductivity, \textbf{c} Lorenz number evolution, and \textbf{d} density of state with increasing doping H content.}
\label{fig:testkkr}
\end{figure*}

The thermal conductivity decreases with H content (Fig.~\ref{fig:transport} b). At the innermost outer core conditions, 0.45 wt\% H (solid deep blue) suppresses the thermal conductivity of hcp phase iron by $\sim$13\%.
DMFT calculations give smaller values than DFT for the thermal conductivity at core conditions. For example, the thermal conductivities of the solid $hcp$ phase at inner core conditions (open blue circles) from DMFT calculations are  $\sim$30 $Wm^{-1}K^{-1}$ lower than predicted by DFT for the same hydrogen content (solid blue circles). Similarly, for the solid phase at outermost outer core conditions, DMFT results (open light green circles)are about $\sim$15 $Wm^{-1}K^{-1}$ lower than those from DFT (solid light green circles). For fluid iron at outermost outer core conditions (solid orange), the effects of H is slightly weaker, especially for DMFT calculations. Unlike the saturation of electrical resistivity in liquid Fe-H alloys, the thermal conductivity keeps increasing with higher temperatures. We obtain thermal conductivity values of $\sim$105 and $\sim$190 $Wm^{-1}K^{-1}$ for the outer and inner core with 0.45 wt\% H content according to DFT calculations, which are respectively about $\sim$13\% and $\sim$8\% lower than the corresponding values in pure iron.

In some previous studies\cite{gomiElectricalResistivityThermal2015,ohtaExperimentalDeterminationElectrical2016}, the thermal conductivity of iron under extreme conditions was obtained  from the electrical resistivity by applying the Wiedemann-Franz law: $\kappa=LT\sigma$, where thermal and electrical conductivity ($\kappa$ and $\sigma$) are associated by temperature T and Lorenz number $L$, with an ideal Lorenz number $L_0=2.44\times10^{-8}W\Omega K^{-2}$. Based on our FPMD+DFT/DMFT results, we compute the Lorenz number $L$ of the Fe-H alloy. The Lorenz number is estimated to be $\sim$(1.95-2.2)$\times10^{-8}W\Omega K^{-2}$ (Fig. \ref{fig:transport}c), which is $\sim$(10-20)\% lower than the ideal value. Overall, we found that H impurities do not alter the Lorenz number significantly. The solid phases have larger values than that of fluid phases at same density of Fe-H alloy and the Lorenz number decreases with higher temperatures. We obtain a Lorenz number of 2.05-2.13$\times10^{-8} W\Omega K^{-2}$ for Fe$_{99.55}$H$_{0.45}$ alloy at the outermost and innermost outer core conditions of Earth. Our results give a  Lorenz number of 2.19 $\times10^{-8}W\Omega K^{-2}$ for pure \textit{hcp} iron at outermost core condition of Earth's, which is consistent with previous DFT computations\cite{zhangReconciliationExperimentsTheory2020}. All the above results pertain to the \(hcp\) Fe-H alloy. For comparison, we also calculated the electrical transport properties of the \(bcc\) Fe-H alloy at the same volume under Earth's inner core conditions (Fig. \ref{fig:testkkr}). The \(bcc\) phase exhibits a distinct trend compared to the \(hcp\) phase. With increasing H content, the electrical resistivity slightly decreases to 53 \(\mu\Omega \cdot \text{cm}\), and the thermal conductivity marginally increases to 246 \(W \cdot \text{m}^{-1} \cdot \text{K}^{-1}\). Examining the density of states (DOS) for the \(bcc\) Fe-H alloy with varying H content (Fig. \ref{fig:testkkr}d), we observed a reduction in the DOS intensity at the Fermi level as H doping increases. This behavior is entirely opposite to that observed in the \(hcp\) phase, where H doping leads to an increase in DOS intensity at the Fermi level.

\begin{figure}[ht]
\centering
\includegraphics[width=0.45\textwidth]{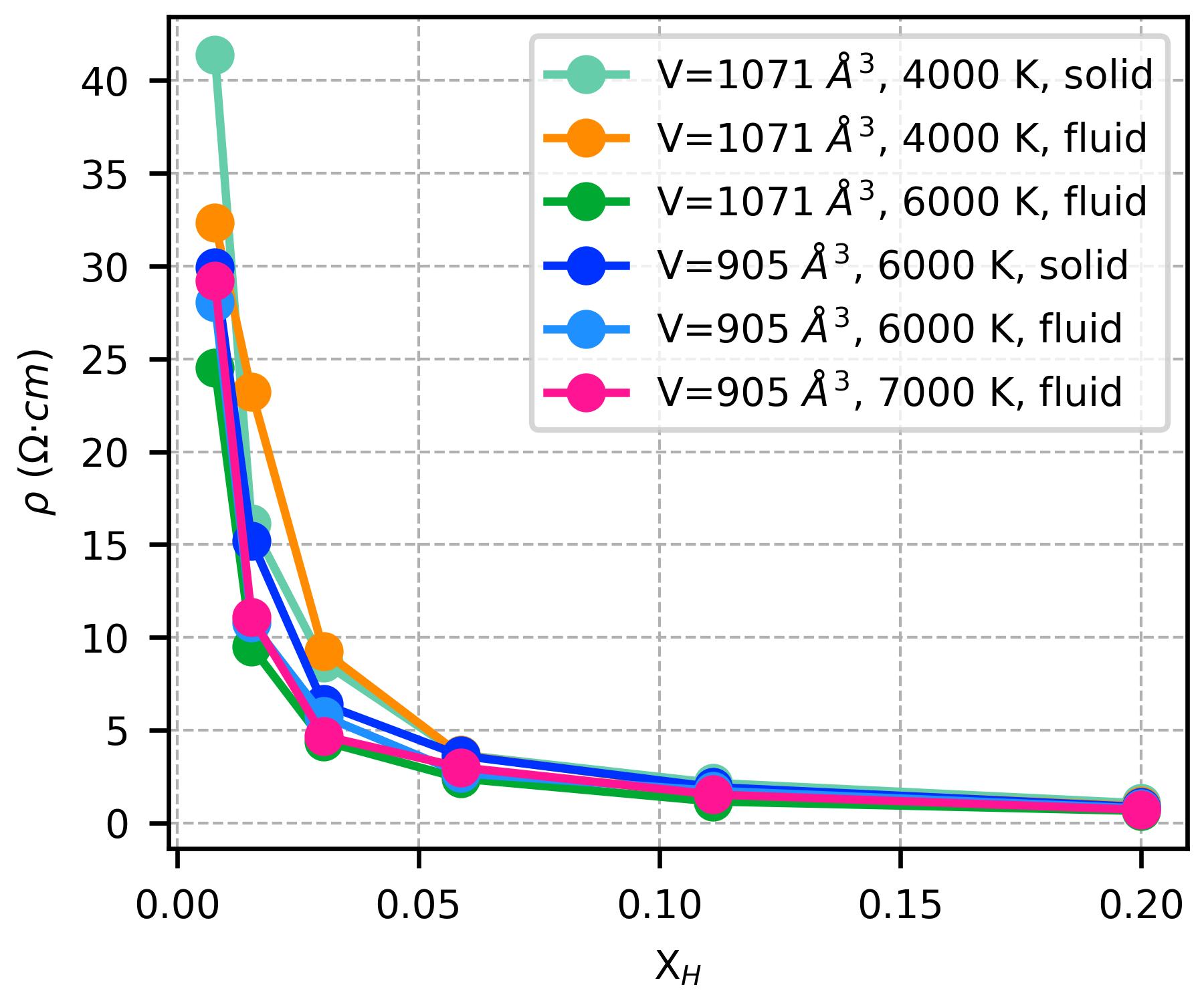}
\caption{\textbf{Ionic resistivity contribution by H diffusivity} volution in two fixed volumes at different temperatures with doping H content by Nernst-Einstein equation. Noting that H diffuses freely in all phases of Fe-H alloy.}
\label{fig:nerest}
\end{figure}

\section{Discussion}
We focused on the electronic contributions to electrical resistivity and thermal conductivity. Previous research has extensively examined the diffusivity of superionic iron-light element alloys at approximately 5000 K, showing slightly higher diffusivity (\(\sim 20\%\) higher) compared to that of Fe-H alloys under Earth's inner core conditions \cite{heSuperionicIronAlloys2022a}. Using these diffusivities, the ionic resistivity contribution from H can be roughly estimated through the Nernst-Einstein equation.
\begin{equation}
    \sigma_{NE}=\frac{e^2q^2_HN_HD_H}{\Omega k_BT},
    \label{eq:nernst}
\end{equation}
where N$_H$ and D$_H$ are the number of hydrogen atoms and their diffusivity, respectively. We roughly set a classical model of charge conduction where hydrogen ions carry an integer charge whose magnitudes equal their formal oxidation numbers q$_H$=1.

As the H content increases, the ionic resistivity of H decreases in all simulations (Fig. \ref{fig:nerest}), as \(\rho\) (\(= 10^8 / \sigma_{NE}\)) strongly depends on \(N_H\). The ionic conductivity is approximately three orders of magnitude lower than the electronic conductivity. The total resistivity is calculated using the parallel resistor formula \cite{wiesmannSimpleModelCharacterizing1977}, and the contribution of ionic resistivity is found to be negligible.
\begin{equation}
    1/\rho=1/\rho_{el}+1/\rho_{ion}
\end{equation}

The interest in determining the transport properties of iron and iron-light element alloys, such as thermal conductivity, is mainly driven by thermal evolution modeling. Directly measuring the thermal properties of these materials at planetary core conditions is challenging, and extrapolations from lower pressure and temperature conditions are not always accurate\cite{ohtaExperimentalDeterminationElectrical2016,suehiroHightemperatureElectricalResistivity2019}. Our calculations indicate a breakdown of the ideal Wiedemann-Franz law in Fe-H alloys, meaning thermal conductivity cannot be indirectly derived from electrical resistivity. Notably, we observed saturation of electrical resistivity in the liquid phase of Fe-H alloy under core conditions, and thermal conductivity increases with temperature. Assuming the Earth's core cools from an early fluid state, the fluid inner core would experience a $\sim$13\% decrease in thermal conductivity from 7000 K to 6000 K, followed by an increase upon crystallization.

Several computational studies have reported evidence suggesting the stability of \(bcc\) iron under Earth's outer core conditions \cite{luoDynamicalStabilityBody2010,belonoshkoStabilizationBodycentredCubic2017b}. However, no experimental X-ray data have been published to confirm the stability of \(bcc\) iron at Earth's core conditions. Recent experiments indicate the stability of the \(hcp\) phase \cite{krausMeasuringMeltingCurve2022}, and simulations suggest that the stability region for the \(bcc\) phase is very narrow, existing only near iron's melting curve. Alloying, such as incorporating silicon, may help stabilize the \(bcc\) phase, particularly at silicon concentrations exceeding \(\sim 10 \, \text{wt}\%\) \cite{fischerPhaseRelationsFe2013}.

Earth’s outer core contains approximately 8 wt\% light elements, such as silicon (Si), oxygen (O), sulfur (S), and carbon (C), along with about 5 wt\% nickel (Ni)\cite{li14ExperimentalConstraints2007,feiThermalEquationState2016}. Recent studies suggest that each weight percent of these light elements can reduce thermal conductivity by 2\%–4\% near core-mantle boundary (CMB) conditions, according to recent calculations and high-pressure experiments\cite{dekokerElectricalResistivityThermal2012,seagleElectricalThermalTransport2013,suehiroInfluenceSulfurElectrical2017,williamsThermalConductivityEarth2018,wagleResistivitySaturationLiquid2019}. Our simulations have only included 0.45 wt\% hydrogen (H) in iron, which reduces thermal conductivity by 8-13\% at the Earth's core conditions than that of the pure iron. The difference in thermal conductivity is significant but reasonable, because hydrogen has a much smaller mass than other light elements. The exact amount of hydrogen in the core is unknowable, but it has a rather minor effect on  transport properties. Since such experiments are unavailable and exceedingly difficult, this is a fortunately result. We also find that melting has only a minor effect on electronic transport in high pressure iron. This is important as there are no experiments on resistivity or thermal conductivity in molten iron. 

\section{Acknowledgments}
\begin{acknowledgments}
This work is supported by US National Science Foundation CSEDI grant EAR-1901813 and the Carnegie Institution for Science. We gratefully acknowledges supercomputer support from Resnick High Performance Computing Center. The authors gratefully acknowledge the Gauss Centre for Supercomputing e.V. (\href{http://www.gauss-centre.eu/}{www.gauss-centre.eu}) for funding this project by providing computing time on the GCS Supercomputer SuperMUC-NG at Leibniz Supercomputing Centre (\href{http://www.lrz.de/}{www.lrz.de}). 
\end{acknowledgments}

\nocite{*}
\bibliography{main.bib}

\newpage
\onecolumngrid
\newpage

\begin{center}
    {Supplementary Materials for \textbf{Electrical resistivity, thermal conductivity, and viscosity of Fe-H alloys at Earth’s core conditions}}
\end{center}

\setcounter{figure}{0}
\counterwithin{figure}{section}
\renewcommand{\thefigure}{S\arabic{figure}}
\setcounter{table}{0}
\counterwithin{table}{section}
\renewcommand{\thetable}{S\arabic{table}}

\maketitle

\onecolumngrid
\section{Simulation details}

\begin{figure*}[ht]
\centering
\includegraphics[width=0.7\textwidth]{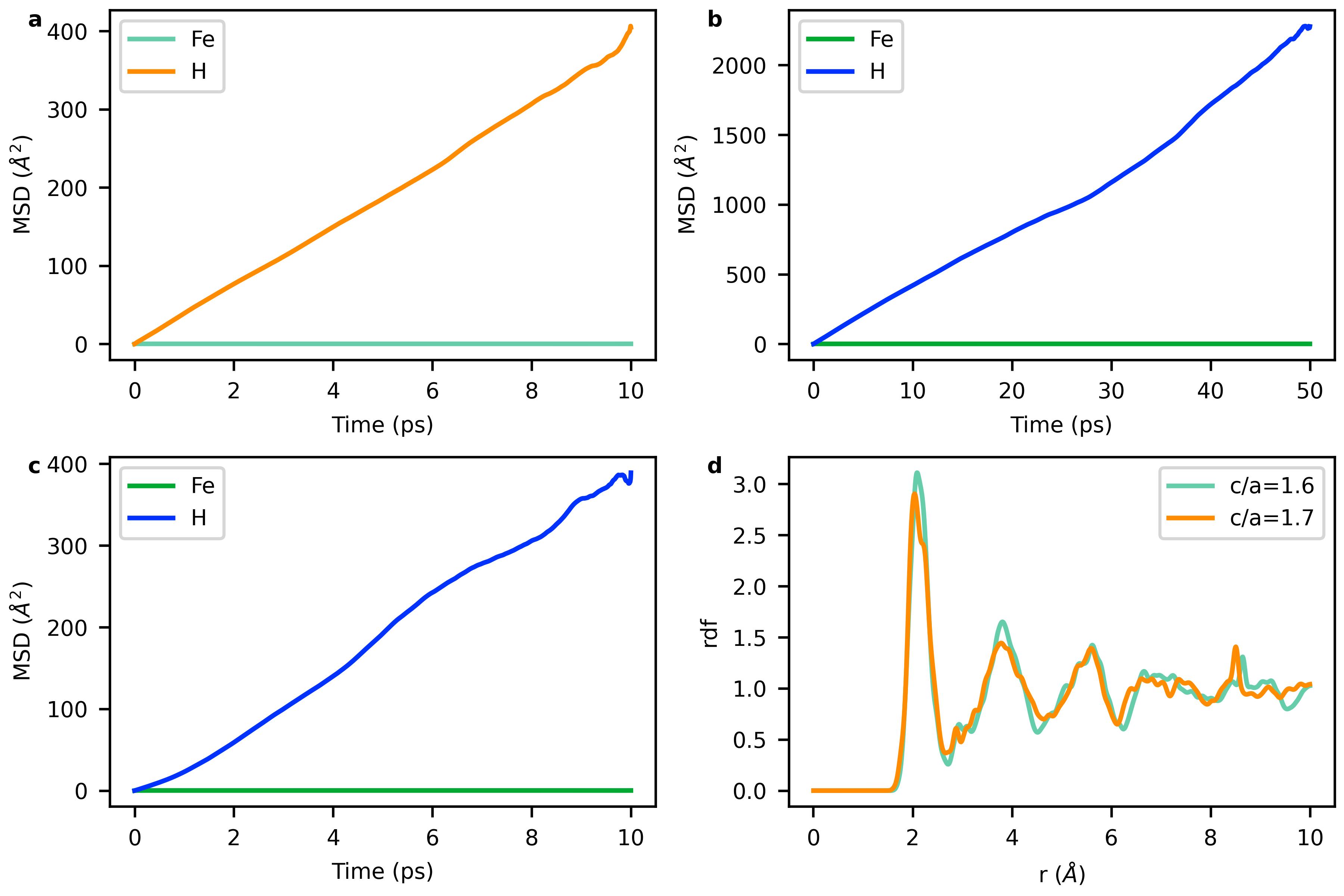}
\caption{\textbf{Test of molecular dynamic simulation of solid phase Fe$_{128}$H$_{32}$ alloy in the volume of 905 $\AA^3$ at 6000 K.} We plotted mean square displacement for Fe and H separately. \textbf{a}~12 ps simulation with time step of 1 fs, and we choose 8 valence electron pesudopotential. \textbf{b}~52 ps simulation with time step of 1 fs, and we choose 8 valence electron pesudopotential \textbf{c}~ 12 ps simulation with time step of 0.25 fs, and we choose 16 valence electron pesudopotential. \textbf{d}~radial distribution functions of hcp phase with c/a ratio of 1.615 and 1.7. We derived the diffusivity from first half of MSD to avoid the increasing error from convolution time.}
\label{fig:mdtest}
\end{figure*}

In order to check our results from Quantum Espresso , we performed some first-principles molecular dynamics (FPMD) simulations with the Vienna ab initio simulation package (VASP)\cite{kresseEfficientIterativeSchemes1996}, using the projector augmented-wave (PAW) method. For VASP, we used 3d$^7$4s$^1$ and 1s$^1$ as valence electrons for Fe and H, and used the generalized gradient approximation (GGA)\cite{blochlProjectorAugmentedwaveMethod1994} in the Perdew-Burke-Ernzerhof (PBE)\cite{perdewGeneralizedGradientApproximation1996a} exchange correlation functional. We also used a 16 valence electron pesudopotential to test these VASP results (Fig. \ref{fig:mdtest}), which was also previously tested for super high pressure region\cite{gonzalez-cataldoInitioDeterminationIron2023} For VASP, we adopted the canonical NVT ensemble using a Nose-Hoover thermostat to perform FPMD simulations in a supercell with 128 atoms Fe and doping H number of 0, 1, 2, 4, 8, 16, 32, with $\Gamma$-centered k-points sampling and a cutoff energy of 600 eV were adopted to ensure energy convergence of better than 10$^{-5}$. At V=1071 $\AA^3$, simulations were conducted at 4000 K to represent the solid/fluid iron phase and at 6000 K for the fluid iron phase. At V=905 $\AA^3$, simulations were conducted at 6000 K to represent the solid/fluid iron phase and at 7000 K for the fluid iron phase. We fixed the c/a ratio to 1.615. Each simulation lasts for 12 ps with a time step of 1 fs, and we allowed the first 2 ps for equilibration. 

We tested our results with regard to simulation time, time step, pseudopotential, and c/a ratios. The radial distribution function of solid phase Fe$_{128}$H$_{32}$ alloy with different c/a ratio for a volume of 705 $\AA^3$ at 6000 K are almost identical (Fig. \ref{fig:mdtest}). The mean square displacement (MSD) also shows only minor differences.

\begin{figure*}[ht]
\centering
\includegraphics[width=0.7\textwidth]{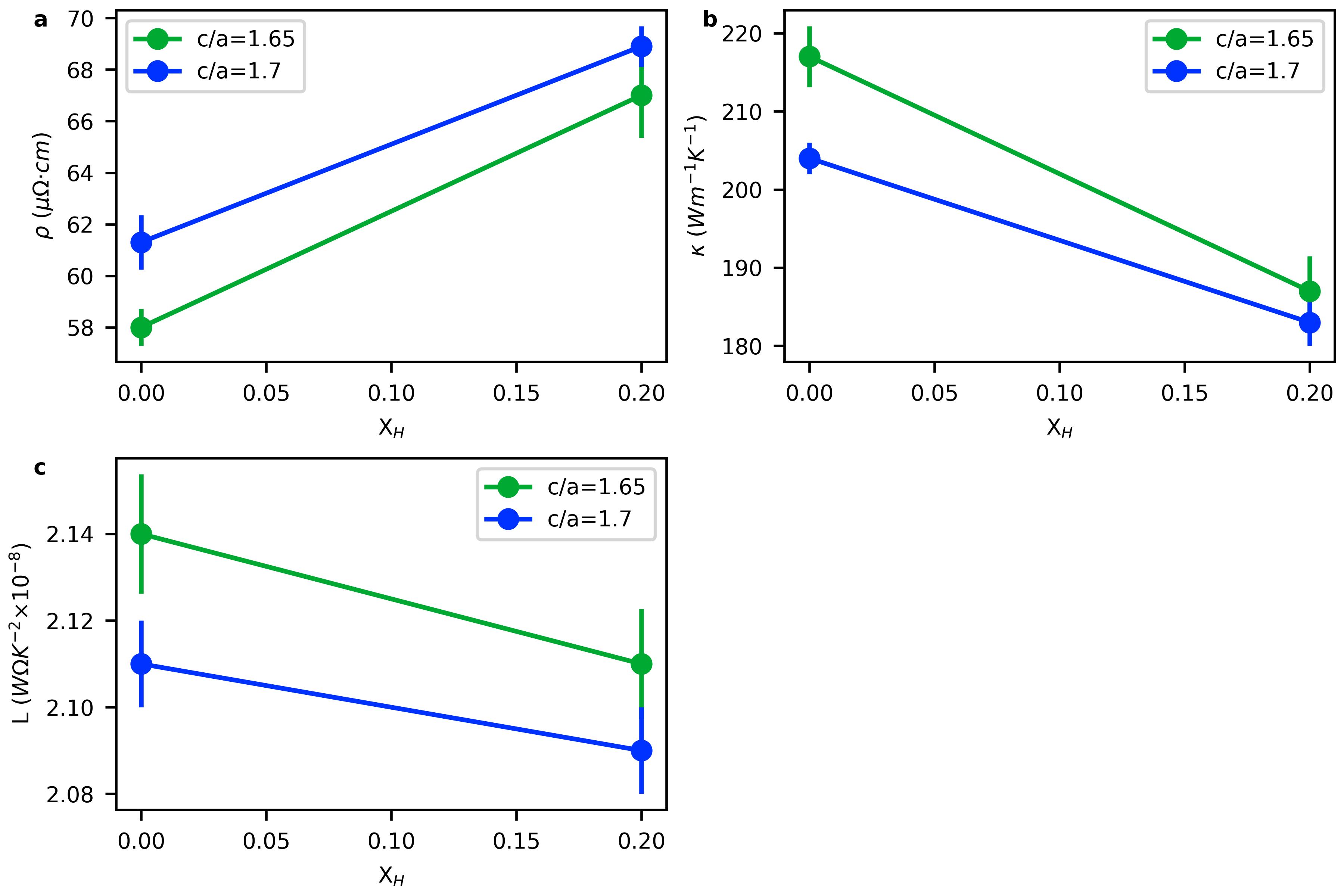}
\caption{\textbf{Electrical transport properties of solid phase Fe-H alloy at 6000 K in the fixed volume of 905 $\AA^3$ for different c/a ratio in KKR.}  \textbf{a} Electrical resistivity, \textbf{b} thermal conductivity and \textbf{c} Lorenz number evolution with different compositions.}
\label{fig:testca}
\end{figure*}

To test convergence of the KKR simulations, we varied nl and c/a ratio, and found only small effects on transport properties (Fig.\ref{fig:testca}). Steinle-Neuman \textit{et al}\cite{steinle2001} suggested that c/a was computed equal to about 1.7 at core conditions. We also compared how electrical properties affect by c/a ration in hcp phase Fe-H alloy below. We can find both electrical resistivities and thermal conductivities of hcp phase Fe-H alloy with c/a ratio constrained to 1.7 keep the same inclination with that of c/a=1.615 case. Compared with c/a=1.615 case, the electrical resistivities are $\sim$2-3 $\mu\Omega\cdot cm$ lower and the thermal conductivities are  $\sim$5-13 $Wm^{-1}K^{-1}$ higher in c/a=1.7 case (Fig.\ref {fig:testca} a and b). We also compared radial distribution functions of these two kinds of MD trajectories with c/a ratio of 1.615 and 1.7 as below. The rdf peaks are very similar (Fig.\ref{fig:mdtest}d).

\section{Supplemental Figures}

\begin{figure}[ht]
\centering
\includegraphics[width=0.8\textwidth]{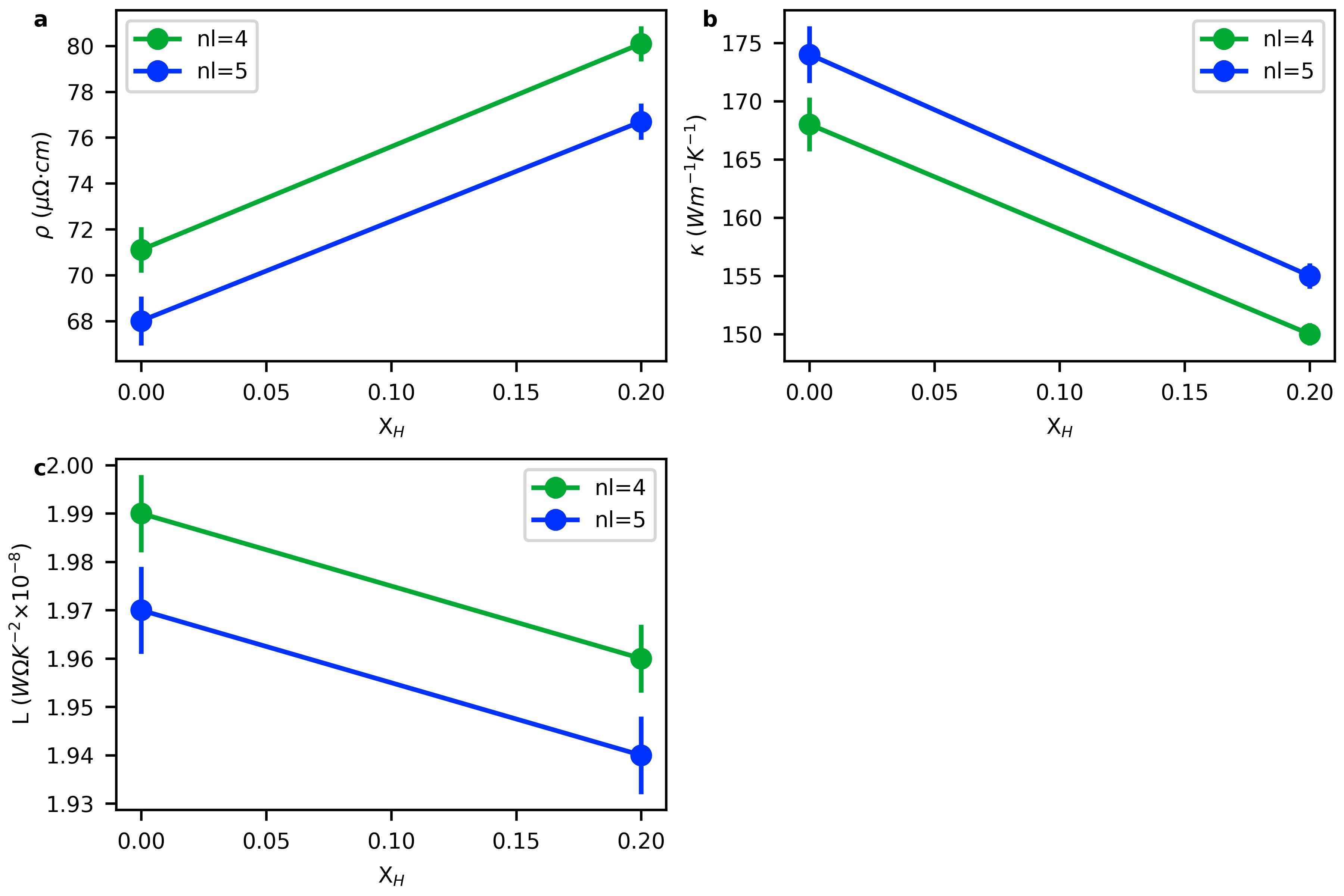}
\caption{\textbf{Electrical transport properties of fluid Fe-H alloy at 6000 K in the fixed volume of 1071 $\AA^3$ for different angular momentum cutoffs in KKR.}  \textbf{a} Electrical resistivity, \textbf{b} thermal conductivity and \textbf{c} Lorenz number evolution with different maximum angular momentum lmax of 3 and 4, corresponding to nl=4 (green) and nl=5 (blue).}
\label{fig:testnl}
\end{figure}

\end{document}